\documentclass[letterpaper]{article}
\pdfoutput=1
\usepackage{hyperref}
\usepackage[english]{babel}
\usepackage[square,sort,comma,numbers]{natbib}
\usepackage{placeins}
\usepackage{tabu}
\usepackage{longtable}
\usepackage{etoolbox}
\usepackage[margin=1in]{geometry}

\hypersetup{colorlinks=true,linkcolor=blue,citecolor=blue,filecolor=blue,urlcolor=blue,
pdftitle={Case Management: a data set of definitions},
pdfauthor={Mike A. Marin, and Matheus Hauder},
pdfsubject={Computer Science, Information systems},
pdfkeywords={Case Handling, Case Management, Case Management Definition}}

\title{Case Management: a data set of definitions}
\author{Mike A. Marin$^{1,2}$, and Matheus Hauder$^{3}$ \\
$^{1}$ IBM Analytics Group\\
Costa Mesa, California, USA\\
$^{2}$ University of South Africa (UNISA)\\
Pretoria, South Africa \\
\texttt{mmarin@acm.org}\\
$^{3}$ Technische Universit\"{a}t M\"{u}nchen (TUM)\\
Boltzmannstr. 3, 85748 Garching bei M\"{u}nchen, Germany\\
\texttt{matheus.hauder@tum.de}}

\begin{document}
\maketitle

\begin{abstract}
Knowledge-intensive processes (KiPs) are becoming increasingly important for organizations with the rise of the knowledge society. Due to their unpredictable and emergent characteristic worklfow management solutions are not suitable to support KiPs. Various case management related approaches have been proposed by researchers and practitioners to support characteristics of KiPs. In this paper we provide a comprehensive list of definitions available on case management, e.g. case handling, adaptive case management, dynamic case management, production case management. For every definition we present the explicit definition, paragraphs that better describe and summarize the case management approach, or extracted sequences that define the term in the referenced publication. All of these definitions are compared against characteristics of KiPs in order to get about understanding of the domain.
\end{abstract}

%\layout{}

\tableofcontents
%\listoffigures
%\listoftables
\pagebreak
\newpage

\section*{Introduction}
\addcontentsline{toc}{section}{Introduction}
We used Google Scholar, IEEExplore, ACM Digital Library, Google Search, and the library of our research institutions using the terms "case management" and "knowledge-intensive process" and trying to identify the first papers that provided definition. Papers in the legal or health care fields were excluded. 
Three types of definitions were extracted, 
\begin{description}
\item[Explicit definitions.] If the paper had an explicit definition that was used.
\item[Paragraphs.] Papers without formal definitions, the paragraph that better described and summarized case management was used as the definition.
\item[Extracted sentences.] Papers without a paragraph that could be used as definition, sentences that defined the term were extracted.
\end{description}
The resulting definitions are presented in the next section.
For each definition, we break it down in what seems to be the main sentences or ideas, and in a set of concepts or components. We also compared the definition with the C1 to C8 KiP characteristics identified by Ciccio \textit{et al.} \cite{DiCiccio2014}.

\section*{Definitions}
\addcontentsline{toc}{section}{Definitions}

\begin{longtabu}{p{16.5cm}}  %\hline
\centering
%%%%%%%%%%%%%%%%%%%%%% BEGIN TABLE %%%%%%%%%%%%%%%%%%%%%%%%%%%%%%%%%%%%%%%%%%%%%
%\begin{table}
%\centering
\begin{tabular}{|p{1.2cm}|p{5cm}|p{9.23cm}|} \hline
1994 
\addcontentsline{toc}{subsection}{1994  Case Management Role by Davenport and Nohria \cite{Davenport1994}} 
& Case Management Role & Davenport and Nohria \cite{Davenport1994} \\\hline
\end{tabular}
\begin{tabular}{|p{16.3cm}|} \hline
The \textbf{case manager role} represents a break with the conventional approach to the division of work. Individuals or small teams perform a series of tasks from beginning to end, often with the help of information systems that reach through the organization. Case managers provide a way to increase organizational efficiency, timeliness, and customer satisfaction. 
 \\\hline
\end{tabular}
\begin{tabular}{|p{9.874cm}|p{6cm}|} \hline
\footnotesize
\textbf{Breakdown:}
\begin{itemize} \setlength\itemsep{-2px}
\item break with the conventional approach to the division of work
\item Individuals or small teams perform a series of tasks from beginning to end
\item a way to increase organizational efficiency, timeliness, and customer satisfaction
\end{itemize}
 &
 \footnotesize
\textbf{components:}
\begin{itemize} \setlength\itemsep{-2px}
\item individuals 
\item information systems
\item small teams 
\item tasks 
\item work 
\end{itemize}\\\hline
\end{tabular}
\begin{tabular}{|c|c|p{2cm}|p{12.4cm}|} \hline
C1 & $\circ$  & \footnotesize Knowledge-driven & \scriptsize The status and availability of data and knowledge objects drive human decision making and directly influence the flow of process actions and events. \\\hline
C2 &  $\circ$ & \footnotesize Collaboration-oriented & \scriptsize Process creation, management and execution occurs in a collaborative multi-user environment, where human-centered and process-related knowledge is co-created, shared and transferred by and among process participants with different roles. \\\hline
C3 &   & \footnotesize Unpredictable & \scriptsize The exact activity, event and knowledge flow depends on situation- and context-specific elements that may not be known a priori, may change during process execution, and may vary over different process cases.\\\hline
C4 &   & \footnotesize Emergent & \scriptsize The actual course of actions gradually emerges during process execution and is determined step by step, when more information is available.\\\hline
C5 &   & \footnotesize Goal-oriented & \scriptsize The process evolves through a series of intermediate goals or milestones to be achieved. \\\hline
C6 &   & \footnotesize Event-driven & \scriptsize Process progression is affected by the occurrence of different kinds of events that influence knowledge workers’ decision making. \\\hline
C7 &   & \footnotesize Constraint- and rule-driven & \scriptsize Process participants may be influenced by or may have to comply with constraints and rules that drive actions performance and decision making. \\\hline
C8 &   & \footnotesize Non-repeatable & \scriptsize The process instance undertaken to deal with a specific case or situation is hardly repeatable, i.e., different executions of the process vary from one another.\\\hline
\end{tabular}
%\caption{1994, Case Management Role by Davenport and Nohria \cite{Davenport1994}}
%\label{table:Davenport1994}
%\end{table}
%%%%%%%%%%%%%%%%%%%%%% END TABLE %%%%%%%%%%%%%%%%%%%%%%%%%%%%%%%%%%%%%%%%%%%%%
~ \\
%~ \\ \hline
%~ \\
%%%%%%%%%%%%%%%%%%%%%% BEGIN TABLE %%%%%%%%%%%%%%%%%%%%%%%%%%%%%%%%%%%%%%%%%%%%%
%\begin{table}
%\centering
\begin{tabular}{|p{1.2cm}|p{5cm}|p{9.23cm}|} \hline
2005 
\addcontentsline{toc}{subsection}{ 2005  Case Handling by van der Aalst \textit{et al.} \cite{Aalst2005}}
& Case Handling & van der Aalst \textit{et al.} \cite{Aalst2005} \\\hline
\end{tabular}
\begin{tabular}{|p{16.3cm}|} \hline
Case handling is a new paradigm for supporting flexible and knowledge intensive business processes. It is strongly based on data as the typical product of these processes. Unlike workflow management, which uses predefined process control structures to determine what should be done during a workflow process, case handling focuses on what can be done to achieve a business goal. In case handling, the knowledge worker in charge of a particular case actively decides on how the goal of that case is reached, and the role of a case handling system is assisting rather than guiding her in doing so.

...

The central concept for case handling is the case and not the activities or the routing. The case is the “product” which is manufactured, and at any time workers should be aware of this context. Examples of cases are the evaluation of a job application, the verdict on a traffic violation, the outcome of a tax assessment, and the ruling for an insurance claim. 

To handle a case, activities need to be executed. Activities are logical units of work.  Many workflow management systems impose the so-called ACID properties on activities. This means that an activity is considered to be atomic and either carried out completely or not at all. Case handling uses a less rigid notion. Activities are simply chunks of work which are recognized by workers, e.g., like filling out an electronic form. As a rule-of-thumb, activities are separated by points where a transfer of work from one worker to another is likely or possible. Please note that activities separated by points of ‘work transfer’ can be non-atomic, e.g., the activity ‘book business trip’ may include tasks such as ‘book flight’, ‘book hotel’, etc. 

Clearly activities are related and cases follow typical patterns. A process is the recipe for handling cases of a given type. In many workflow management systems, the specification of a process fixes the routing of cases along activities, and workers have hardly any insight in the whole. As a result exceptions are difficult to handle because they require unparalleled deviations from the standard recipe.
 \\\hline
\end{tabular}
\begin{tabular}{|p{9.874cm}|p{6cm}|} \hline
\footnotesize
\textbf{Breakdown:}
\begin{itemize} \setlength\itemsep{-2px}
\item a new paradigm
\item support flexible and knowledge intensive business processes
\item based on data
\item data is the typical product of these processes
\item focus on a business goal
\item knowledge worker decides on how the goal is reached
\item role of a case handling system is assisting rather than guiding her
\item central concept is the case and not the activities or the routing
\item case is the “product” which is manufactured
\item  workers should be aware of this context
\item  Activities are logical units of work
\item Case handling uses a less rigid notion of activities
\item  Activities are simply chunks of work which are recognized by workers
\item  activities are separated by points where a transfer of work from one worker to another is likely or possible
\item activities separated by points of ‘work transfer’ can be non-atomic
\item  activities are related and cases follow typical patterns
\item A process is the recipe for handling cases of a given type
\item  exceptions are easy to handle 
\end{itemize}
 &
 \footnotesize
\textbf{components:}
\begin{itemize} \setlength\itemsep{-2px}
\item activities
\item assisting 
\item based on data
\item business goal
\item case
\item case handling
\item chunks of work
\item exceptions
\item flexible
\item goal
\item knowledge intensive
\item knowledge worker
\item logical units of work
\item manufactured
\item non-atomic
\item recipe
\item routing
\item transfer of work
\item workers
\end{itemize}\\\hline
\end{tabular}
\begin{tabular}{|c|c|p{2cm}|p{12.4cm}|} \hline
C1 & $\bullet$  & \footnotesize Knowledge-driven & \scriptsize The status and availability of data and knowledge objects drive human decision making and directly influence the flow of process actions and events. \\\hline
C2 &   & \footnotesize Collaboration-oriented & \scriptsize Process creation, management and execution occurs in a collaborative multi-user environment, where human-centered and process-related knowledge is co-created, shared and transferred by and among process participants with different roles. \\\hline
C3 & $\circ$  & \footnotesize Unpredictable & \scriptsize The exact activity, event and knowledge flow depends on situation- and context-specific elements that may not be known a priori, may change during process execution, and may vary over different process cases.\\\hline
C4 & $\bullet$  & \footnotesize Emergent & \scriptsize The actual course of actions gradually emerges during process execution and is determined step by step, when more information is available.\\\hline
C5 & $\bullet$  & Goal-oriented & The process evolves through a series of intermediate goals or milestones to be achieved. \\\hline
C6 &   & \footnotesize Event-driven & \scriptsize Process progression is affected by the occurrence of different kinds of events that influence knowledge workers’ decision making. \\\hline
C7 &   & \footnotesize Constraint- and rule-driven & \scriptsize Process participants may be influenced by or may have to comply with constraints and rules that drive actions performance and decision making. \\\hline
C8 &   & \footnotesize Non-repeatable & \scriptsize The process instance undertaken to deal with a specific case or situation is hardly repeatable, i.e., different executions of the process vary from one another.\\\hline
\end{tabular}
%\caption{CAPTION}
%\label{table:Types}
%\end{table}
%%%%%%%%%%%%%%%%%%%%%% END TABLE %%%%%%%%%%%%%%%%%%%%%%%%%%%%%%%%%%%%%%%%%%%%%
~ \\
%~ \\ \hline
%~ \\
%%%%%%%%%%%%%%%%%%%%%% BEGIN TABLE %%%%%%%%%%%%%%%%%%%%%%%%%%%%%%%%%%%%%%%%%%%%%
%\begin{table}
%\centering
\begin{tabular}{|p{1.2cm}|p{5cm}|p{9.23cm}|} \hline
2006 
\addcontentsline{toc}{subsection}{ 2006  Case Management by Kaan \textit{et al.} \cite{Kaan2006}}
 & Case Management & Kaan \textit{et al.} \cite{Kaan2006} \\\hline
\end{tabular}
\begin{tabular}{|p{16.3cm}|} \hline
\textbf{No formal definition. Described in terms of the technology}:

What actually happens within the execution of a task for a particular case is of no concern to the WfMS. Aside from monitoring changes to case attribute values that may be of influence on further routing, the tasks in the workflow process are “black boxes” as far as the WfMS is concerned.
...
case management is not advocated as a substitute for workflow management but as a harmonious extension.
...
we propose the use of Workflow Management (including its methods, techniques and tools) on the flow of control level and introduce the case management paradigm on the work contents level.
Within such a scope, it can be assumed that workers will receive work items from a
WfMS, which are instantiations of particular tasks for particular cases. ... To complete a work item, a set of activities needs to be performed. A central element in case management is that the characteristics of the case determine what activities are considered to be relevant during case processing. This relevance is considered as the context of activities and expressed in terms of conditions on the case attributes.
...
case management concept presupposes a very fine-grained view on the activities that people perform within the context of carrying out a work item.
...
case management concept as we describe it in this paper, the AM supports the use of time frames in which activities are relevant, the starting of activities in parallel, the definition of a repository of plug-ins to be started by the user and an ad-hoc taskbar for starting common functionality from anywhere in the process.
...
Case management in particular introduced the benefit of increased flexibility and uniformity in user interface presentation
...
case management, ... address[es]... the issues of flexibility and awareness.
...
Case management (CM) takes out the coordination logic within single tasks from the generic application. Instead, a generic system is available that offers both support and flexibility to the performer for executing a work item
 \\\hline
\end{tabular}
\begin{tabular}{|p{9.874cm}|p{6cm}|} \hline
\tiny % \footnotesize
\textbf{Breakdown:}
\begin{itemize} \setlength\itemsep{-2px}
\item What actually happens within the execution of a task for a particular case is of no concern to the WfMS
\item  tasks in the workflow process are “black boxes” as far as the WfMS is concerned
\item case management is not advocated as a substitute for workflow management but as a harmonious extension.
\item case management paradigm on the work contents level.
\item workers will receive work items from a WfMS
\item To complete a work item a set of activities needs to be performed.
\item  A central element in case management is that the characteristics of the case determine what activities are considered to be relevant during case processing.
\item the context of activities and expressed in terms of conditions on the case attributes.
\item case management concept presupposes a very fine-grained view on the activities that people perform within the context of carrying out a work item.
\item the Activity Manage supports the use of time frames in which activities are relevant, the starting of activities in parallel, the definition of a repository of plug-ins to be started by the user and an ad-hoc taskbar for starting common functionality from anywhere in the process.
\item introduced the benefit of increased flexibility and uniformity in user interface presentation
\item addresses the issues of flexibility and awareness.
\item Case management takes out the coordination logic within single tasks from the generic application.
\item  a generic system is available that offers both support and flexibility to the performer for executing a work item
\end{itemize}
 &
\tiny % \footnotesize
\textbf{components:}
\begin{itemize} \setlength\itemsep{-3px}
\item activities
\item activity manager
\item ad-hoc taskbar
\item awareness.
\item black boxes
\item case
\item case attributes
\item case management
\item case processing
\item conditions
\item context
\item control level
\item coordination logic
\item flexibility
\item generic application
\item parallel
\item people
\item performer
\item plug-ins
\item repository
\item routing
\item single tasks
\item support
\item task
\item time frames
\item uniformity
\item user
\item user interface
\item work contents level
\item work item
\item workers
\item workflow
\end{itemize}
\\\hline
\end{tabular}
\begin{tabular}{|c|c|p{2cm}|p{12.4cm}|} \hline
C1 &   & \footnotesize Knowledge-driven & \scriptsize The status and availability of data and knowledge objects drive human decision making and directly influence the flow of process actions and events. \\\hline
C2 &   & \footnotesize Collaboration-oriented & \scriptsize Process creation, management and execution occurs in a collaborative multi-user environment, where human-centered and process-related knowledge is co-created, shared and transferred by and among process participants with different roles. \\\hline
C3 & $-$  & \footnotesize Unpredictable & \scriptsize The exact activity, event and knowledge flow depends on situation- and context-specific elements that may not be known a priori, may change during process execution, and may vary over different process cases.\\\hline
C4 & $-$  & \footnotesize Emergent & \scriptsize The actual course of actions gradually emerges during process execution and is determined step by step, when more information is available.\\\hline
C5 &   & Goal-oriented & The process evolves through a series of intermediate goals or milestones to be achieved. \\\hline
C6 &  $-$ & \footnotesize Event-driven & \scriptsize Process progression is affected by the occurrence of different kinds of events that influence knowledge workers’ decision making. \\\hline
C7 &   & \footnotesize Constraint- and rule-driven & \scriptsize Process participants may be influenced by or may have to comply with constraints and rules that drive actions performance and decision making. \\\hline
C8 & $-$  & \footnotesize Non-repeatable & \scriptsize The process instance undertaken to deal with a specific case or situation is hardly repeatable, i.e., different executions of the process vary from one another.\\\hline
\end{tabular}
%\caption{2006, Case Management by Kaan \textit{et al.} \cite{Kaan2006}}
%\label{table:Kaan2006}
%\end{table}
%%%%%%%%%%%%%%%%%%%%%% END TABLE %%%%%%%%%%%%%%%%%%%%%%%%%%%%%%%%%%%%%%%%%%%%%
~ \\
%~ \\ \hline
%~ \\
%%%%%%%%%%%%%%%%%%%%%% BEGIN TABLE %%%%%%%%%%%%%%%%%%%%%%%%%%%%%%%%%%%%%%%%%%%%%
%\begin{table}
%\centering
\begin{tabular}{|p{1.2cm}|p{5cm}|p{9.23cm}|} \hline
2008 
\addcontentsline{toc}{subsection}{ 2008  Case Management Work by Kerremans \cite{Kerremans2008}}
 & Case Management Work & Kerremans \cite{Kerremans2008} \\\hline
\end{tabular}
\begin{tabular}{|p{16.3cm}|} \hline
\textbf{Case management work} is collaborative and nondeterministic, meaning it has one or more points where different continuations are possible, and it departs from traditional structured, sequential predefined processes. Case management work depends more on human decision making and content than other processes do.
 \\\hline
\end{tabular}
\begin{tabular}{|p{9.874cm}|p{6cm}|} \hline
\footnotesize
\textbf{Breakdown:}
\begin{itemize} \setlength\itemsep{-2px}
\item collaborative and nondeterministic (it has one or more points where different continuations are possible)
\item departs from traditional structured, sequential predefined processes
\item depends more on human decision making and content than other processes do
\end{itemize}
 &
 \footnotesize
\textbf{components:}
\begin{itemize} \setlength\itemsep{-2px}
\item content
\item human decision
\item process
\item work
\end{itemize}
\\\hline
\end{tabular}
\begin{tabular}{|c|c|p{2cm}|p{12.4cm}|} \hline
C1 & $\circ$  & \footnotesize Knowledge-driven & \scriptsize The status and availability of data and knowledge objects drive human decision making and directly influence the flow of process actions and events. \\\hline
C2 & $\bullet$  & \footnotesize Collaboration-oriented & \scriptsize Process creation, management and execution occurs in a collaborative multi-user environment, where human-centered and process-related knowledge is co-created, shared and transferred by and among process participants with different roles. \\\hline
C3 & $\circ$  & \footnotesize Unpredictable & \scriptsize The exact activity, event and knowledge flow depends on situation- and context-specific elements that may not be known a priori, may change during process execution, and may vary over different process cases.\\\hline
C4 & $\bullet$  & \footnotesize Emergent & \scriptsize The actual course of actions gradually emerges during process execution and is determined step by step, when more information is available.\\\hline
C5 &   & \footnotesize Goal-oriented & \scriptsize The process evolves through a series of intermediate goals or milestones to be achieved. \\\hline
C6 &   & \footnotesize Event-driven & \scriptsize Process progression is affected by the occurrence of different kinds of events that influence knowledge workers’ decision making. \\\hline
C7 &   & \footnotesize Constraint- and rule-driven & \scriptsize Process participants may be influenced by or may have to comply with constraints and rules that drive actions performance and decision making. \\\hline
C8 & $\circ$  & \footnotesize Non-repeatable & \scriptsize The process instance undertaken to deal with a specific case or situation is hardly repeatable, i.e., different executions of the process vary from one another.\\\hline
\end{tabular}
%\caption{2008, Case Management Work by Kerremans \cite{Kerremans2008}}
%\label{table:Kerremans2008}
%\end{table}
%%%%%%%%%%%%%%%%%%%%%% END TABLE %%%%%%%%%%%%%%%%%%%%%%%%%%%%%%%%%%%%%%%%%%%%%
~ \\
~ \\
~ \\
~ \\
~ \\
~ \\
~ \\
~ \\
%~ \\ \hline
%~ \\
%%%%%%%%%%%%%%%%%%%%%% BEGIN TABLE %%%%%%%%%%%%%%%%%%%%%%%%%%%%%%%%%%%%%%%%%%%%%
%\begin{table}
%\centering
\begin{tabular}{|p{1.2cm}|p{5cm}|p{9.23cm}|} \hline
 2009 
\addcontentsline{toc}{subsection}{ 2009 Dynamic Case Management by Clair \textit{et al.} \cite{Clair2009}}
 & Dynamic Case Management & Clair \textit{et al.} \cite{Clair2009} \\\hline
\end{tabular}
\begin{tabular}{|p{16.3cm}|} \hline
\textbf{Dynamic Case Management}: A highly structured, but also collaborative, dynamic, and information-intensive process that is driven by outside events and requires incremental and progressive responses from the business domain handling the case. Examples of case folders include a patient record, a lawsuit, an insurance claim, or a contract, and the case folder would include all the documents, data, collaboration artifacts, policies, rules, analytics, and other information needed to process and manage the case.
 \\\hline
\end{tabular}
\begin{tabular}{|p{9.874cm}|p{6cm}|} \hline
\footnotesize
\textbf{Breakdown:}
\begin{itemize} \setlength\itemsep{-2px}
\item highly structured, but also collaborative, dynamic, and information-intensive process
\item driven by outside events
\item requires incremental and progressive responses from the business domain handling the case
\item case folder would include all the documents, data, collaboration artifacts, policies, rules, analytics, and other information needed to process and manage the case
\end{itemize}
 &
 \footnotesize
\textbf{components:}
\begin{itemize} \setlength\itemsep{-2px}
\item analytics
\item artifacts
\item case
\item case folder
\item data
\item documents
\item events
\item information
\item policies
\item process
\item rules
\end{itemize}
\\\hline
\end{tabular}
\begin{tabular}{|c|c|p{2cm}|p{12.4cm}|} \hline
C1 & $\circ$  & \footnotesize Knowledge-driven & \scriptsize The status and availability of data and knowledge objects drive human decision making and directly influence the flow of process actions and events. \\\hline
C2 & $\bullet$  & \footnotesize Collaboration-oriented & \scriptsize Process creation, management and execution occurs in a collaborative multi-user environment, where human-centered and process-related knowledge is co-created, shared and transferred by and among process participants with different roles. \\\hline
C3 &   & \footnotesize Unpredictable & \scriptsize The exact activity, event and knowledge flow depends on situation- and context-specific elements that may not be known a priori, may change during process execution, and may vary over different process cases.\\\hline
C4 &   & \footnotesize Emergent & \scriptsize The actual course of actions gradually emerges during process execution and is determined step by step, when more information is available.\\\hline
C5 &   & \footnotesize Goal-oriented & \scriptsize The process evolves through a series of intermediate goals or milestones to be achieved. \\\hline
C6 & $\bullet$  & \footnotesize Event-driven & \scriptsize Process progression is affected by the occurrence of different kinds of events that influence knowledge workers’ decision making. \\\hline
C7 & $\circ$  & \footnotesize Constraint \& rule-driven & \scriptsize Process participants may be influenced by or may have to comply with constraints and rules that drive actions performance and decision making. \\\hline
C8 &   & \footnotesize Non-repeatable & \scriptsize The process instance undertaken to deal with a specific case or situation is hardly repeatable, i.e., different executions of the process vary from one another.\\\hline
\end{tabular}
%\caption{2009, Dynamic Case Management by Clair \textit{et al.} \cite{Clair2009}}
%\label{table:Clair2009}
%\end{table}
%%%%%%%%%%%%%%%%%%%%%% END TABLE %%%%%%%%%%%%%%%%%%%%%%%%%%%%%%%%%%%%%%%%%%%%%
~ \\
%~ \\ \hline
%~ \\
%%%%%%%%%%%%%%%%%%%%%% BEGIN TABLE %%%%%%%%%%%%%%%%%%%%%%%%%%%%%%%%%%%%%%%%%%%%%
%\begin{table}
%\centering
\begin{tabular}{|p{1.2cm}|p{5cm}|p{9.23cm}|} \hline
 2009 
\addcontentsline{toc}{subsection}{2009  Case Management by White \cite{White2009}}
 & Case Management & White \cite{White2009} \\\hline
\end{tabular}
\begin{tabular}{|p{16.3cm}|} \hline
\textbf{Case Management} is the management of long-lived collaborative processes that coordinate knowledge, content, correspondence and resources to progress a case to achieve a particular goal; where the path of execution cannot be predetermined in advance of execution; where human judgment is required to determine how the end goal can be achieved; and where the state of a case can be altered by external out-of-band events.
 \\\hline
\end{tabular}
\begin{tabular}{|p{9.874cm}|p{6cm}|} \hline
\footnotesize
\textbf{Breakdown:}
\begin{itemize} \setlength\itemsep{-2px}
\item management of long-lived collaborative processes
\item coordinate knowledge, content, correspondence and resources to progress a case to achieve a particular goal
\item the path of execution cannot be predetermined in advance
\item where human judgment is required to determine how the end goal can be achieved
\item where the state of a case can be altered by external out-of-band events
\end{itemize}
 &
 \footnotesize
\textbf{components:}
\begin{itemize} \setlength\itemsep{-2px}
\item case
\item collaborative processes
\item content
\item correspondence
\item events
\item goal
\item human judgment
\item knowledge
\item resources
\item state of a case
\end{itemize}
\\\hline
\end{tabular}
\begin{tabular}{|c|c|p{2cm}|p{12.4cm}|} \hline
C1 &  $\bullet$ & \footnotesize Knowledge-driven & \scriptsize The status and availability of data and knowledge objects drive human decision making and directly influence the flow of process actions and events. \\\hline
C2 & $\bullet$  & \footnotesize Collaboration-oriented & \scriptsize Process creation, management and execution occurs in a collaborative multi-user environment, where human-centered and process-related knowledge is co-created, shared and transferred by and among process participants with different roles. \\\hline
C3 &  $\circ$ & \footnotesize Unpredictable & \scriptsize The exact activity, event and knowledge flow depends on situation- and context-specific elements that may not be known a priori, may change during process execution, and may vary over different process cases.\\\hline
C4 & $\circ$  & \footnotesize Emergent & \scriptsize The actual course of actions gradually emerges during process execution and is determined step by step, when more information is available.\\\hline
C5 & $\bullet$  & \footnotesize Goal-oriented & \scriptsize The process evolves through a series of intermediate goals or milestones to be achieved. \\\hline
C6 & $\bullet$  & \footnotesize Event-driven & \scriptsize Process progression is affected by the occurrence of different kinds of events that influence knowledge workers’ decision making. \\\hline
C7 &   & \footnotesize Constraint \& rule-driven & \scriptsize Process participants may be influenced by or may have to comply with constraints and rules that drive actions performance and decision making. \\\hline
C8 & $\circ$  & \footnotesize Non-repeatable & \scriptsize The process instance undertaken to deal with a specific case or situation is hardly repeatable, i.e., different executions of the process vary from one another.\\\hline
\end{tabular}
%\caption{2009, Case Management by White \cite{White2009}}
%\label{table:White2009}
%\end{table}
%%%%%%%%%%%%%%%%%%%%%% END TABLE %%%%%%%%%%%%%%%%%%%%%%%%%%%%%%%%%%%%%%%%%%%%%
~ \\
~ \\
~ \\
~ \\
~ \\
~ \\
~ \\
~ \\
~ \\
~ \\
~ \\
~ \\
~ \\
~ \\
~ \\
~ \\
~ \\
~ \\
~ \\
~ \\
~ \\
~ \\
~ \\
~ \\
~ \\
~ \\
~ \\
\addcontentsline{toc}{subsection}{2010 Mastering the Unpredictable \cite{Swenson2010}}
\textbf{2010 Mastering the Unpredictable \cite{Swenson2010}} \\
The following five definitions were extracted from Swenson \cite{Swenson2010} \\
~ \\
~ \\
%~ \\ \hline
%~ \\
%%%%%%%%%%%%%%%%%%%%%% BEGIN TABLE %%%%%%%%%%%%%%%%%%%%%%%%%%%%%%%%%%%%%%%%%%%%%
%\begin{table}
%\centering
\begin{tabular}{|p{1.2cm}|p{5cm}|p{9.23cm}|} \hline
2010 
\addcontentsline{toc}{subsubsection}{2010 Case Management by McCauley \cite{McCauley2010}}
 & Case Management & McCauley \cite{McCauley2010} \\\hline
\end{tabular}
\begin{tabular}{|p{16.3cm}|} \hline
\textbf{Case management} is the management of long-lived collaborative processes that require coordination of knowledge, content, correspondence, and resources to achieve an objective or goal. The path of execution cannot be predefined. Human judgment is required in determining how to proceed, and the state of a case can be affected by external events.
 \\\hline
\end{tabular}
\begin{tabular}{|p{9.874cm}|p{6cm}|} \hline
\footnotesize
\textbf{Breakdown:}
\begin{itemize} \setlength\itemsep{-2px}
\item management of long-lived collaborative processes
\item require coordination of knowledge, content, correspondence, and resources to achieve an objective or goal
\item The path of execution cannot be predefined
\item Human judgment is required in determining how to proceed
\item the state of a case can be affected by external events
\end{itemize}
 &
 \footnotesize
\textbf{components:}
\begin{itemize} \setlength\itemsep{-2px}
\item case
\item collaborative process
\item content
\item correspondence
\item events
\item goal
\item human judgment
\item knowledge
\item resources
\item state
\end{itemize}
\\\hline
\end{tabular}
\begin{tabular}{|c|c|p{2cm}|p{12.4cm}|} \hline
C1 & $\bullet$  & \footnotesize Knowledge-driven & \scriptsize The status and availability of data and knowledge objects drive human decision making and directly influence the flow of process actions and events. \\\hline
C2 & $\bullet$  & \footnotesize Collaboration-oriented & \scriptsize Process creation, management and execution occurs in a collaborative multi-user environment, where human-centered and process-related knowledge is co-created, shared and transferred by and among process participants with different roles. \\\hline
C3 & $\bullet$  & \footnotesize Unpredictable & \scriptsize The exact activity, event and knowledge flow depends on situation- and context-specific elements that may not be known a priori, may change during process execution, and may vary over different process cases.\\\hline
C4 & $\bullet$ & \footnotesize Emergent & \scriptsize The actual course of actions gradually emerges during process execution and is determined step by step, when more information is available.\\\hline
C5 & $\bullet$  & \footnotesize Goal-oriented & \scriptsize The process evolves through a series of intermediate goals or milestones to be achieved. \\\hline
C6 & $\bullet$  & \footnotesize Event-driven & \scriptsize Process progression is affected by the occurrence of different kinds of events that influence knowledge workers’ decision making. \\\hline
C7 &   & \footnotesize Constraint \& rule-driven & \scriptsize Process participants may be influenced by or may have to comply with constraints and rules that drive actions performance and decision making. \\\hline
C8 & $\circ$  & \footnotesize Non-repeatable & \scriptsize The process instance undertaken to deal with a specific case or situation is hardly repeatable, i.e., different executions of the process vary from one another.\\\hline
\end{tabular}
%\caption{2010, Case Management by McCauley \cite{McCauley2010}}
%\label{table:McCauley2010}
%\end{table}
%%%%%%%%%%%%%%%%%%%%%% END TABLE %%%%%%%%%%%%%%%%%%%%%%%%%%%%%%%%%%%%%%%%%%%%%
~ \\
%~ \\ \hline
%~ \\
%%%%%%%%%%%%%%%%%%%%%% BEGIN TABLE %%%%%%%%%%%%%%%%%%%%%%%%%%%%%%%%%%%%%%%%%%%%%
%\begin{table}
%\centering
\begin{tabular}{|p{1.2cm}|p{5cm}|p{9.23cm}|} \hline
 2010 
 \addcontentsline{toc}{subsubsection}{2010 Case Management by de Man \textit{et al.} \cite{deMan2010}}
 & Case Management & de Man \textit{et al.} \cite{deMan2010} \\\hline
\end{tabular}
\begin{tabular}{|p{16.3cm}|} \hline
\textbf{Case management} is a method or practice of coordinating work by organizing all of the relevant information into one place. A case is a specific situation to take care of. The representation of the case in the case management system is thought of as holding all of the information and processes, and it coordinates communications necessary to accomplish the goal for that particular situation. The case becomes the focal point for assessing the situation, initiating activities and processes, as well as keeping a history record of what has transpired.
 \\\hline
\end{tabular}
\begin{tabular}{|p{9.874cm}|p{6cm}|} \hline
\footnotesize
\textbf{Breakdown:}
\begin{itemize} \setlength\itemsep{-2px}
\item method or practice of coordinating work
\item organizing all of the relevant information into one place
\item A case is a specific situation to take care of
\item holding all of the information and processes
\item coordinates communications necessary to accomplish the goal for that particular situation
\item becomes the focal point for assessing the situation, initiating activities and processes
\item keeping a history record of what has transpired
\end{itemize}
 &
 \footnotesize
\textbf{components:}
\begin{itemize} \setlength\itemsep{-2px}
\item activities
\item case
\item case management system
\item communications
\item goal
\item history
\item information
\item one place (focal place)
\item particular situation
\item processes
\item specific situation
\item work
\end{itemize}
\\\hline
\end{tabular}
\begin{tabular}{|c|c|p{2cm}|p{12.4cm}|} \hline
C1 &   & \footnotesize Knowledge-driven & \scriptsize The status and availability of data and knowledge objects drive human decision making and directly influence the flow of process actions and events. \\\hline
C2 &   & \footnotesize Collaboration-oriented & \scriptsize Process creation, management and execution occurs in a collaborative multi-user environment, where human-centered and process-related knowledge is co-created, shared and transferred by and among process participants with different roles. \\\hline
C3 &   & \footnotesize Unpredictable & \scriptsize The exact activity, event and knowledge flow depends on situation- and context-specific elements that may not be known a priori, may change during process execution, and may vary over different process cases.\\\hline
C4 &  & \footnotesize Emergent & \scriptsize The actual course of actions gradually emerges during process execution and is determined step by step, when more information is available.\\\hline
C5 & $\bullet$  & \footnotesize Goal-oriented & \scriptsize The process evolves through a series of intermediate goals or milestones to be achieved. \\\hline
C6 &   & \footnotesize Event-driven & \scriptsize Process progression is affected by the occurrence of different kinds of events that influence knowledge workers’ decision making. \\\hline
C7 &   & \footnotesize Constraint \& rule-driven & \scriptsize Process participants may be influenced by or may have to comply with constraints and rules that drive actions performance and decision making. \\\hline
C8 &   & \footnotesize Non-repeatable & \scriptsize The process instance undertaken to deal with a specific case or situation is hardly repeatable, i.e., different executions of the process vary from one another.\\\hline
\end{tabular}
%\caption{2010, Case Management by de Man \textit{et al.} \cite{deMan2010}}
%\label{table:deMan2010}
%\end{table}
%%%%%%%%%%%%%%%%%%%%%% END TABLE %%%%%%%%%%%%%%%%%%%%%%%%%%%%%%%%%%%%%%%%%%%%%
~ \\
%~ \\ \hline
%~ \\
%%%%%%%%%%%%%%%%%%%%%% BEGIN TABLE %%%%%%%%%%%%%%%%%%%%%%%%%%%%%%%%%%%%%%%%%%%%%
%\begin{table}
%\centering
\begin{tabular}{|p{1.2cm}|p{5cm}|p{9.23cm}|} \hline
 2010 
 \addcontentsline{toc}{subsubsection}{2010  Case Management in Glossary \cite{Swenson2010,Swenson2015}}
 & Case Management & Glossary \cite{Swenson2010,Swenson2015} \\\hline
\end{tabular}
\begin{tabular}{|p{16.3cm}|} \hline
\textbf{Case management}: A method or practice of coordinating work by organizing all of the relevant information into one place called a case. The case becomes the focal point for assessing the situation, initiating activities and processes, as well as keeping a history record of what has transpired. 
 \\\hline
\end{tabular}
\begin{tabular}{|p{9.874cm}|p{6cm}|} \hline
\footnotesize
\textbf{Breakdown:}
\begin{itemize} \setlength\itemsep{-2px}
\item method or practice of coordinating work
\item organizing all of the relevant information into one place
\item the focal point for assessing the situation, initiating activities and processes
\item keeping a history record of what has transpired
\end{itemize}
 &
 \footnotesize
\textbf{components:}
\begin{itemize} \setlength\itemsep{-2px}
\item activities
\item case
\item history
\item information
\item processes
\item situation
\item work
\end{itemize}
\\\hline
\end{tabular}
\begin{tabular}{|c|c|p{2cm}|p{12.4cm}|} \hline
C1 &   & \footnotesize Knowledge-driven & \scriptsize The status and availability of data and knowledge objects drive human decision making and directly influence the flow of process actions and events. \\\hline
C2 &   & \footnotesize Collaboration-oriented & \scriptsize Process creation, management and execution occurs in a collaborative multi-user environment, where human-centered and process-related knowledge is co-created, shared and transferred by and among process participants with different roles. \\\hline
C3 &   & \footnotesize Unpredictable & \scriptsize The exact activity, event and knowledge flow depends on situation- and context-specific elements that may not be known a priori, may change during process execution, and may vary over different process cases.\\\hline
C4 &  & \footnotesize Emergent & \scriptsize The actual course of actions gradually emerges during process execution and is determined step by step, when more information is available.\\\hline
C5 &   & \footnotesize Goal-oriented & \scriptsize The process evolves through a series of intermediate goals or milestones to be achieved. \\\hline
C6 &   & \footnotesize Event-driven & \scriptsize Process progression is affected by the occurrence of different kinds of events that influence knowledge workers’ decision making. \\\hline
C7 &   & \footnotesize Constraint \& rule-driven & \scriptsize Process participants may be influenced by or may have to comply with constraints and rules that drive actions performance and decision making. \\\hline
C8 &   & \footnotesize Non-repeatable & \scriptsize The process instance undertaken to deal with a specific case or situation is hardly repeatable, i.e., different executions of the process vary from one another.\\\hline
\end{tabular}
%\caption{2010, Case Management in Glossary of \cite{Swenson2010}}
%\label{table:Swenson2010}
%\end{table}
%%%%%%%%%%%%%%%%%%%%%% END TABLE %%%%%%%%%%%%%%%%%%%%%%%%%%%%%%%%%%%%%%%%%%%%%
~ \\
%~ \\ \hline
%~ \\
%%%%%%%%%%%%%%%%%%%%%% BEGIN TABLE %%%%%%%%%%%%%%%%%%%%%%%%%%%%%%%%%%%%%%%%%%%%%
%\begin{table}
%\centering
\begin{tabular}{|p{1.2cm}|p{5cm}|p{9.23cm}|} \hline
 2010 
\addcontentsline{toc}{subsubsection}{2010  Adaptive Case Management by Swenson \cite{Swenson2010a}}
 & Adaptive Case Management & Swenson \cite{Swenson2010a} \\\hline
\end{tabular}
\begin{tabular}{|p{16.3cm}|} \hline
\textbf{Adaptive case management}: A productive system that deploys not only the organization and process structure, but it becomes the systems of record for the business data entities and content involved. All processes are completely transparent, as per access authorization, and fully auditable. It enables nontechnical business users in virtual organizations to seamlessly create/consolidate structured and unstructured processes from basic predefined business entities, content, social interactions, and business rules. It moves the process knowledge gathering from the template analysis/modeling/simulation phase into the process execution phase in the lifecycle. It collects actionable knowledge  without an intermediate analysis phase  based on process patterns created by business users. ACM differs from business process management (BPM) in that the case information is the focus and the thing around which the other artifacts are organized. And it is the case information that persists for the long term.
 \\\hline
\end{tabular}
\begin{tabular}{|p{9.874cm}|p{6cm}|} \hline
\footnotesize
\textbf{Breakdown:}
\begin{itemize} \setlength\itemsep{-2px}
\item A productive system that deploys the organization and process structure
\item becomes the systems of record for the business data entities and content involved
\item processes are completely transparent, as per access authorization, and fully auditable
\item enables nontechnical business users in virtual organizations to seamlessly create/consolidate structured and unstructured processes
\item basic predefined business entities, content, social interactions, and business rules
\item moves the process knowledge gathering from the template analysis/modeling/simulation phase into the process execution phase in the lifecycle
\item collects actionable knowledge  without an intermediate analysis phase  based on process patterns created by business users
\item differs from business process management (BPM) in that the case information is the focus and the thing around which the other artifacts are organized
\item case information  persists for the long term
\end{itemize}
 &
\footnotesize
\textbf{components:}
\begin{itemize} \setlength\itemsep{-2px}
\item access authorization
\item actionable knowledge
\item artifacts
\item auditable
\item business data
\item business entities
\item business rules
\item business users
\item case information
\item content
\item lifecycle
\item organization
\item process execution
\item process knowledge
\item process patterns
\item process structure
\item processes
\item productive system
\item social interactions
\item structured processes 
\item systems of record
\item template analysis
\item template modeling
\item template simulation
\item transparent
\item unstructured processes
\item virtual organizations
\end{itemize}
\\\hline
\end{tabular}
\begin{tabular}{|c|c|p{2cm}|p{12.4cm}|} \hline
C1 &   & \footnotesize Knowledge-driven & \scriptsize The status and availability of data and knowledge objects drive human decision making and directly influence the flow of process actions and events. \\\hline
C2 & $\circ$  & \footnotesize Collaboration-oriented & \scriptsize Process creation, management and execution occurs in a collaborative multi-user environment, where human-centered and process-related knowledge is co-created, shared and transferred by and among process participants with different roles. \\\hline
C3 &   & \footnotesize Unpredictable & \scriptsize The exact activity, event and knowledge flow depends on situation- and context-specific elements that may not be known a priori, may change during process execution, and may vary over different process cases.\\\hline
C4 &  & \footnotesize Emergent & \scriptsize The actual course of actions gradually emerges during process execution and is determined step by step, when more information is available.\\\hline
C5 &   & \footnotesize Goal-oriented & \scriptsize The process evolves through a series of intermediate goals or milestones to be achieved. \\\hline
C6 &   & \footnotesize Event-driven & \scriptsize Process progression is affected by the occurrence of different kinds of events that influence knowledge workers’ decision making. \\\hline
C7 &   & \footnotesize Constraint \& rule-driven & \scriptsize Process participants may be influenced by or may have to comply with constraints and rules that drive actions performance and decision making. \\\hline
C8 &   & \footnotesize Non-repeatable & \scriptsize The process instance undertaken to deal with a specific case or situation is hardly repeatable, i.e., different executions of the process vary from one another.\\\hline
\end{tabular}
%\caption{2010, Adaptive Case Management by Swenson \cite{Swenson2010}}
%\label{table:Swenson2010a}
%\end{table}
%%%%%%%%%%%%%%%%%%%%%% END TABLE %%%%%%%%%%%%%%%%%%%%%%%%%%%%%%%%%%%%%%%%%%%%%
~ \\
%~ \\ \hline
%~ \\
%%%%%%%%%%%%%%%%%%%%%% BEGIN TABLE %%%%%%%%%%%%%%%%%%%%%%%%%%%%%%%%%%%%%%%%%%%%%
%\begin{table}
%\centering
\begin{tabular}{|p{1.2cm}|p{5cm}|p{9.23cm}|} \hline
 2010 
\addcontentsline{toc}{subsubsection}{2010  Adaptive Case Management  by Palmer \cite{Palmer2010}}
 & Adaptive Case Management  & Palmer \cite{Palmer2010} \\\hline
\end{tabular}
\begin{tabular}{|p{16.3cm}|} \hline
\textbf{Adaptive case management}: Systems that are able to support decision making and data capture while providing the freedom for knowledge workers to apply their own understanding and subject matter expertise to respond to unique or changing circumstances within the business environment.
 \\\hline
\end{tabular}
\begin{tabular}{|p{9.874cm}|p{6cm}|} \hline
\footnotesize
\textbf{Breakdown:}
\begin{itemize} \setlength\itemsep{-2px}
\item able to support decision making
\item able to support  data capture
\item provide the freedom for knowledge workers to apply their own understanding and subject matter expertise to respond to unique or changing circumstances within the business environment
\end{itemize}
 &
 \footnotesize
\textbf{components:}
\begin{itemize} \setlength\itemsep{-2px}
\item business environment
\item data capture
\item decision making
\item knowledge workers
\item subject matter expertise
\item systems
\item understanding
\end{itemize}
\\\hline
\end{tabular}
\begin{tabular}{|c|c|p{2cm}|p{12.4cm}|} \hline
C1 & $\circ$  & \footnotesize Knowledge-driven & \scriptsize The status and availability of data and knowledge objects drive human decision making and directly influence the flow of process actions and events. \\\hline
C2 &   & \footnotesize Collaboration-oriented & \scriptsize Process creation, management and execution occurs in a collaborative multi-user environment, where human-centered and process-related knowledge is co-created, shared and transferred by and among process participants with different roles. \\\hline
C3 &   & \footnotesize Unpredictable & \scriptsize The exact activity, event and knowledge flow depends on situation- and context-specific elements that may not be known a priori, may change during process execution, and may vary over different process cases.\\\hline
C4 &  & \footnotesize Emergent & \scriptsize The actual course of actions gradually emerges during process execution and is determined step by step, when more information is available.\\\hline
C5 &   & \footnotesize Goal-oriented & \scriptsize The process evolves through a series of intermediate goals or milestones to be achieved. \\\hline
C6 &   & \footnotesize Event-driven & \scriptsize Process progression is affected by the occurrence of different kinds of events that influence knowledge workers’ decision making. \\\hline
C7 &   & \footnotesize Constraint \& rule-driven & \scriptsize Process participants may be influenced by or may have to comply with constraints and rules that drive actions performance and decision making. \\\hline
C8 &   & \footnotesize Non-repeatable & \scriptsize The process instance undertaken to deal with a specific case or situation is hardly repeatable, i.e., different executions of the process vary from one another.\\\hline
\end{tabular}
%\caption{2010, Adaptive Case Management  by Palmer \cite{Palmer2010}}
%\label{table:Palmer2010}
%\end{table}
%%%%%%%%%%%%%%%%%%%%%% END TABLE %%%%%%%%%%%%%%%%%%%%%%%%%%%%%%%%%%%%%%%%%%%%%
~ \\
~ \\
~ \\
~ \\
~ \\
~ \\
~ \\
~ \\
%~ \\ \hline
%~ \\
%%%%%%%%%%%%%%%%%%%%%% BEGIN TABLE %%%%%%%%%%%%%%%%%%%%%%%%%%%%%%%%%%%%%%%%%%%%%
%\begin{table}
%\centering
\begin{tabular}{|p{1.2cm}|p{5cm}|p{9.23cm}|} \hline
 2011 
\addcontentsline{toc}{subsection}{2011  Emerging Case Management by B\"{o}hringer \cite{Bohringer2011}}
 & Emerging Case Management & B\"{o}hringer \cite{Bohringer2011} \\\hline
\end{tabular}
\begin{tabular}{|p{16.3cm}|} \hline
\textbf{Case Management} is an approach to handling ad-hoc processes. It supports process instances (cases) in providing a collaborative space to store and negotiate case-related activities and artifacts (e.g. information, decisions, workflows) without the need for the ex ante modeling of the whole process.
 \\\hline
\end{tabular}
\begin{tabular}{|p{9.874cm}|p{6cm}|} \hline
\footnotesize
\textbf{Breakdown:}
\begin{itemize} \setlength\itemsep{-2px}
\item approach to handling ad-hoc processes
\item providing a collaborative space
\item store and negotiate case-related activities and artifacts
\item no need for modeling of the whole process
\end{itemize}
 &
 \footnotesize
\textbf{components:}
\begin{itemize} \setlength\itemsep{-2px}
\item activities
\item ad-hoc process
\item artifacts
\item case
\item collaborative space
\item decisions
\item information
\item process instances
\item workflows
\end{itemize}
\\\hline
\end{tabular}
\begin{tabular}{|c|c|p{2cm}|p{12.4cm}|} \hline
C1 & $\circ$  & \footnotesize Knowledge-driven & \scriptsize The status and availability of data and knowledge objects drive human decision making and directly influence the flow of process actions and events. \\\hline
C2 & $\bullet$  & \footnotesize Collaboration-oriented & \scriptsize Process creation, management and execution occurs in a collaborative multi-user environment, where human-centered and process-related knowledge is co-created, shared and transferred by and among process participants with different roles. \\\hline
C3 & $\circ$  & \footnotesize Unpredictable & \scriptsize The exact activity, event and knowledge flow depends on situation- and context-specific elements that may not be known a priori, may change during process execution, and may vary over different process cases.\\\hline
C4 &  & \footnotesize Emergent & \scriptsize The actual course of actions gradually emerges during process execution and is determined step by step, when more information is available.\\\hline
C5 &   & \footnotesize Goal-oriented & \scriptsize The process evolves through a series of intermediate goals or milestones to be achieved. \\\hline
C6 &   & \footnotesize Event-driven & \scriptsize Process progression is affected by the occurrence of different kinds of events that influence knowledge workers’ decision making. \\\hline
C7 &   & \footnotesize Constraint \& rule-driven & \scriptsize Process participants may be influenced by or may have to comply with constraints and rules that drive actions performance and decision making. \\\hline
C8 &   & \footnotesize Non-repeatable & \scriptsize The process instance undertaken to deal with a specific case or situation is hardly repeatable, i.e., different executions of the process vary from one another.\\\hline
\end{tabular}
%\caption{2011, Emerging Case Management by B\"{o}hringer \cite{Bohringer2011}}
%\label{table:Bohringer2011}
%\end{table}
%%%%%%%%%%%%%%%%%%%%%% END TABLE %%%%%%%%%%%%%%%%%%%%%%%%%%%%%%%%%%%%%%%%%%%%%
~ \\
%~ \\ \hline
%~ \\
%%%%%%%%%%%%%%%%%%%%%% BEGIN TABLE %%%%%%%%%%%%%%%%%%%%%%%%%%%%%%%%%%%%%%%%%%%%%
%\begin{table}
%\centering
\begin{tabular}{|p{1.2cm}|p{5cm}|p{9.23cm}|} \hline
 2013 
\addcontentsline{toc}{subsection}{2013  Production Case Management by Swenson \cite{Swenson2013} and Motahari-Nezhad and Swenson \cite{Motahari2013,Swenson2015}}
 & Production Case Management & Swenson \cite{Swenson2013} and Motahari-Nezhad and Swenson \cite{Motahari2013,Swenson2015}\\\hline
\end{tabular}
\begin{tabular}{|p{16.3cm}|} \hline
\textbf{Production Case Management} (PCM) is an approach to supporting knowledge workers which is programmed by specially-trained technical people (programmers) to produce a case management application. That application is deployed for use by knowledge workers to get their work done. The application offers collections of operations that the knowledge worker can select to use or not use depending on the specific needs of the case. A PCM application is used when there is a certain amount of unpredictability in the work, and some flexibility is needed, but necessary actions are regular enough or the volume of work large enough to make identifying and codifying regular patterns valuable. A worker using PCM will be involved in selecting the actions toward the outcome of a particular case, but will not be responsible for the kinds of actions that might be available.
 \\\hline
\end{tabular}
\begin{tabular}{|p{9.874cm}|p{6cm}|} \hline
\footnotesize
\textbf{Breakdown:}
\begin{itemize} \setlength\itemsep{-2px}
\item approach to support knowledge workers
\item programmed by specially-trained technical people (programmers) to produce a case management application. 
\item The application is deployed for use by knowledge workers to get their work done. 
\item The application offers collections of operations that the knowledge worker can select to use or not use depending on the specific needs of the case. 
\item used when there is a certain amount of unpredictability in the work, and some flexibility is needed
\item necessary actions are regular enough or the volume of work large enough to make identifying and codifying regular patterns valuable. 
\item a worker will be involved in selecting the actions toward the outcome of a particular case, 
\item a worker will not be responsible for the kinds of actions that might be available.
\end{itemize}
 &
\footnotesize
\textbf{components:}
\begin{itemize} \setlength\itemsep{-2px}
\item actions
\item application
\item case
\item case management application
\item collections of operations
\item flexibility
\item knowledge workers
\item outcome
\item regular patterns
\item specially-trained technical people (programmers)
\item volume of work
\item work
\end{itemize}
\\\hline
\end{tabular}
\begin{tabular}{|c|c|p{2cm}|p{12.4cm}|} \hline
C1 & $\bullet$  & \footnotesize Knowledge-driven & \scriptsize The status and availability of data and knowledge objects drive human decision making and directly influence the flow of process actions and events. \\\hline
C2 &   & \footnotesize Collaboration-oriented & \scriptsize Process creation, management and execution occurs in a collaborative multi-user environment, where human-centered and process-related knowledge is co-created, shared and transferred by and among process participants with different roles. \\\hline
C3 & $\bullet$  & \footnotesize Unpredictable & \scriptsize The exact activity, event and knowledge flow depends on situation- and context-specific elements that may not be known a priori, may change during process execution, and may vary over different process cases.\\\hline
C4 & $\circ$ & \footnotesize Emergent & \scriptsize The actual course of actions gradually emerges during process execution and is determined step by step, when more information is available.\\\hline
C5 & $\circ$  & \footnotesize Goal-oriented & \scriptsize The process evolves through a series of intermediate goals or milestones to be achieved. \\\hline
C6 &   & \footnotesize Event-driven & \scriptsize Process progression is affected by the occurrence of different kinds of events that influence knowledge workers’ decision making. \\\hline
C7 &   & \footnotesize Constraint \& rule-driven & \scriptsize Process participants may be influenced by or may have to comply with constraints and rules that drive actions performance and decision making. \\\hline
C8 &   & \footnotesize Non-repeatable & \scriptsize The process instance undertaken to deal with a specific case or situation is hardly repeatable, i.e., different executions of the process vary from one another.\\\hline
\end{tabular}
%\caption{2013, Production Case Management by Swenson \cite{Swenson2013} and \cite{Motahari2013,Swenson2015}}
%\label{table:Swenson2013}
%\end{table}
%%%%%%%%%%%%%%%%%%%%%% END TABLE %%%%%%%%%%%%%%%%%%%%%%%%%%%%%%%%%%%%%%%%%%%%%
~ \\
%~ \\ \hline
%~ \\
%%%%%%%%%%%%%%%%%%%%%% BEGIN TABLE %%%%%%%%%%%%%%%%%%%%%%%%%%%%%%%%%%%%%%%%%%%%%
%\begin{table}
%\centering
\begin{tabular}{|p{1.2cm}|p{5cm}|p{9.23cm}|} \hline
 2013 
\addcontentsline{toc}{subsection}{2013 Adaptive Case Management by Swenson \cite{Swenson2013} and Motahari-Nezhad and Swenson \cite{Motahari2013,Swenson2015}}
 & Adaptive Case Management & Swenson \cite{Swenson2013} and Motahari-Nezhad and Swenson \cite{Motahari2013,Swenson2015} \\\hline
\end{tabular}
\begin{tabular}{|p{16.3cm}|} \hline
\textbf{Adaptive Case Management} (ACM) is an approach to support knowledge workers who need the most flexibility to handle their cases. ACM allows the knowledge workers themselves to create and modify all aspects of a case at any time. There is no distinction between design time and run time: the designing and running are done at the same time by the same people. This approach is used by knowledge workers who have unique expertise in an area. They don't have a lot of time to transfer this specific knowledge to a programmer, and it is too expensive to hire a programmer for one-off cases. ACM offers a Do-It-Yourself (DIY) approach to process programming. The worker using ACM is responsible not only for the outcome of a case, but also for how the handling of that kind of case improves over time.
 \\\hline
\end{tabular}
\begin{tabular}{|p{9.874cm}|p{6cm}|} \hline
\footnotesize
\textbf{Breakdown:}
\begin{itemize} \setlength\itemsep{-2px} 
\item approach to support knowledge workers who need the most flexibility to handle their cases
\item allows the knowledge workers themselves to create and modify all aspects of a case at any time. 
\item no distinction between design time and run time
\item the designing and running are done at the same time by the same people
\item used by knowledge workers who have unique expertise in an area
\item knowledge workers don't have a lot of time to transfer this specific knowledge to a programmer
\item it is too expensive to hire a programmer for one-off cases
\item offers a Do-It-Yourself (DIY) approach to process programming
\item The worker using ACM is responsible not only for the outcome of a case, but also for how the handling of that kind of case improves over time.
\end{itemize}
 &
 \footnotesize
\textbf{components:}
\begin{itemize} \setlength\itemsep{-2px}
\item area
\item cases
\item design time
\item do-It-Yourself (DIY)
\item expertise
\item knowledge
\item knowledge workers
\item people
\item process programming
\item programmer
\item run time
\end{itemize}
\\\hline
\end{tabular}
\begin{tabular}{|c|c|p{2cm}|p{12.4cm}|} \hline
C1 & $\bullet$  & \footnotesize Knowledge-driven & \scriptsize The status and availability of data and knowledge objects drive human decision making and directly influence the flow of process actions and events. \\\hline
C2 &   & \footnotesize Collaboration-oriented & \scriptsize Process creation, management and execution occurs in a collaborative multi-user environment, where human-centered and process-related knowledge is co-created, shared and transferred by and among process participants with different roles. \\\hline
C3 & $\bullet$  & \footnotesize Unpredictable & \scriptsize The exact activity, event and knowledge flow depends on situation- and context-specific elements that may not be known a priori, may change during process execution, and may vary over different process cases.\\\hline
C4 & $\bullet$ & \footnotesize Emergent & \scriptsize The actual course of actions gradually emerges during process execution and is determined step by step, when more information is available.\\\hline
C5 & $\circ$ & \footnotesize Goal-oriented & \scriptsize The process evolves through a series of intermediate goals or milestones to be achieved. \\\hline
C6 &   & \footnotesize Event-driven & \scriptsize Process progression is affected by the occurrence of different kinds of events that influence knowledge workers’ decision making. \\\hline
C7 &   & \footnotesize Constraint \& rule-driven & \scriptsize Process participants may be influenced by or may have to comply with constraints and rules that drive actions performance and decision making. \\\hline
C8 & $\circ$  & \footnotesize Non-repeatable & \scriptsize The process instance undertaken to deal with a specific case or situation is hardly repeatable, i.e., different executions of the process vary from one another.\\\hline
\end{tabular}
%\caption{2013, Production Case Management by Swenson \cite{Swenson2013} and \cite{Motahari2013,Swenson2015}}
%\label{table:Swenson2013a}
%\end{table}
%%%%%%%%%%%%%%%%%%%%%% END TABLE %%%%%%%%%%%%%%%%%%%%%%%%%%%%%%%%%%%%%%%%%%%%%
~ \\ %\hline
\end{longtabu}

\FloatBarrier
\pagebreak
\newpage
\addcontentsline{toc}{section}{References}
\bibliographystyle{abbrv}
\bibliography{CMMN-KiP}

\end{document}